\documentclass[11pt]{article}

\usepackage[]{acl}

\usepackage{times}
\usepackage{latexsym}

\usepackage[T1]{fontenc}

\usepackage[utf8]{inputenc}

\usepackage{microtype}

\usepackage{inconsolata}

\usepackage{graphicx}
\usepackage{booktabs}

\usepackage{longtable}
\usepackage{booktabs}
\usepackage{array}
\usepackage{ragged2e}
\usepackage{multirow}
\newcolumntype{P}[1]{>{\RaggedRight\arraybackslash}p{#1}}

\usepackage{tcolorbox}
\usepackage{xcolor}
\usepackage{enumitem}

\usepackage{caption}
\usepackage{subcaption}
\usepackage{blindtext}

\usepackage{listings}
\lstdefinestyle{jsonwrap}{
  basicstyle=\ttfamily\footnotesize,
  breaklines=true,
  breakatwhitespace=false,
  columns=fullflexible,
  keepspaces=true,
  showstringspaces=false,
  frame=none
}

\title{The Creation and Analysis of \\ Government AI
Transparency Statements in Australia}    

\author{
 \textbf{Shidong Pan\textsuperscript{1}\thanks{Shidong Pan completed the work when he was a visiting research scientist at CSIRO's Data61.}},
 \textbf{Haochen Gong\textsuperscript{2}},
 \textbf{Boming Xia\textsuperscript{3}},
 \\
 \textbf{Xiaoyu Sun\textsuperscript{2}},
 \textbf{Xiwei Xu\textsuperscript{1}},
 \textbf{Liming Zhu\textsuperscript{1}}
\\
 \textsuperscript{1}CSIRO,
 \textsuperscript{2}Australian National University,
 \textsuperscript{3}University of Adelaide
\\
 \small{
   \textbf{Correspondence:} \href{xiwei.xu@data61.csiro.au}{xiwei.xu@data61.csiro.au}
 }
}

\begin{document}
\maketitle

\begin{abstract}

Governments increasingly deploy AI in public services, making transparency essential for accountability and public trust. 
Australia's Standard for AI Transparency Statements (AITS) requires government bodies to disclose how AI is used in practice, yet little empirical evidence exists on how these requirements are realised in documents.
This paper presents a government AITS dataset, dubbed AITS-101, and provides one of the first systematic analysis of their content. 
Using stylometric, quantitative, and qualitative document analyses, we examine disclosure coverage, structure, and recurring patterns. 
Our findings reveal substantial variation in AI-related practice disclosure, highlight gaps between policy intent and implementation, and inform the design of more effective public-sector AI transparency standards.

\end{abstract}

\section{Introduction}

The advancement of AI has underscored its capacity to enhance productivity, strengthen analytical and decision-making processes, and support more responsive public services~\citep{wirtz2019artificial, mergel2024implementing, czarnitzki2023artificial, brynjolfsson2025generative}. 
This potential is widely recognised across the international community, where many governments have begun deploying AI in public administration and have introduced governance policies, such as EU AI Act\footnote{\url{https://artificialintelligenceact.eu/}} and the United State’s President Executive Order 14110,\footnote{\url{federalregister.gov/documents/2023/11/01/2023-24283/use-of-artificial-intelligence}} to regulate its responsible use.

Across these regulatory efforts, \textbf{transparency} consistently emerges as a foundational principle. 
Obligations related to transparency serve to inform both affected individuals and relevant oversight bodies regarding the functioning and decision-making processes of AI systems.
In public-sector settings, transparency is particularly critical for maintaining public trust and democratic legitimacy~\cite{lund2025standards, hoang2026esg, weatherall2026aitransparency}.
However, the deployment of AI carries inherent risks that must be carefully managed. 
Ongoing concerns surrounding the ethical and accountable use of AI highlight the need for safeguards to ensure that government adoption of such technologies remains aligned with public expectations~\citep{lu2024responsible}.
Within this regulatory landscape, \underline{\textbf{AI}} \underline{\textbf{T}}ransparency \underline{\textbf{S}}tatements (AITS) have emerged as a concrete policy instrument to operationalise transparency obligations in government settings.

In Australia, the government introduced the Policy for the Responsible Use of AI in Government~\citep{australia-ai-transparency-standard-2025}, initially released in August 2024 and revised in December 2025, outlining a whole-of-government commitment to safe, ethical, and responsible AI use. 
Under this policy, all non-corporate Commonwealth entities (NCEs) are required to publish AITS on their publicly accessible websites. 
These statements must follow the structure and content requirements specified in the Standard for AI Transparency Statements \citep{australia-ai-transparency-standard-2025}, which aims to ensure that government agencies provide clear, consistent, and accessible explanations of how AI systems are used in practice.

In this paper, we present one of the first systematic empirical study of AI Transparency Statements.  
We compile and release \textbf{AITS-101}, a curated dataset of 101 publicly available AITS from Australian government agencies, providing a unique snapshot of how AI transparency requirements are documented and operationalised in practice. 
Using a mixed-methods document analysis approach, we examine these statements from three complementary perspectives. 
First, we conduct a stylometric analysis to characterise their length, structure, and linguistic features. 
Second, we perform a quantitative composition analysis based on a fine-grained annotation taxonomy to assess the coverage and distribution of disclosure AI-related practices.
Third, we carry out a qualitative analysis to identify recurring disclosure patterns and their broader implications for public-sector AI transparency. 

Together, our findings reveal systematic gaps between policy intent and documented disclosure practices, offering empirical evidence to inform the design, evaluation, and evolution of AITS in the public sector.

\section{Related Work}

\subsection{Public Sector AI Transparency}

Public trust is a foundational requirement for effective public-sector governance~\cite{andrews2022trust, matulionyte2023regulating, lund2025standards, hoang2026esg}. 
As governments increasingly deploy AI in decision-making and service delivery, transparency has become a central concern.
Prior research on public-sector AI transparency has largely focused on legal obligations and governance frameworks. 
Legal scholarship examines how existing mechanisms, such as freedom of information regimes, shape access to information about government AI use, while highlighting structural limitations related to proprietary and institutional constraints~\cite{olsen2024right, zhang2025right}. 
Conceptual work also investigates how AI transparency is framed in governmental discourse and proposes high-level frameworks and best practices for public-sector disclosure~\cite{sebastiao2025ai}.
Despite this growing literature, empirical analyses of how AI transparency is implemented in practice, particularly through disclosure documents, remain limited.

Weatherall et al., \cite{weatherall2026aitransparency} provide the closest empirical study of Australian AITS, focusing primarily on whether Commonwealth entities publish accessible and meaningful statements and identifying gaps in availability, discoverability, and compliance quality.
Our work complements this by treating AITS as a corpus of public-facing disclosure documents and conducting systematic stylometric, content, and qualitative analyses of what agencies disclose and how they communicate AI use in practice.


%

\subsection{Document Analysis of Transparency Statements}

AITS belong to a broader class of disclosure documents designed to enhance digital transparency by communicating complex technical and organisational practices to the public. 
The most extensively studied example is the privacy policy, which serves as a primary mechanism for disclosing data handling practices in digital systems~\cite{chang2018role, pan2024trap, schaub2025grand, pan2023toward}.
Prior studies on privacy policies have established robust methodological foundations, including stylometric analysis~\cite{pan2024trap}, longitudinal evolution analysis~\cite{amos2021privacy, tao2025longitudinal}, and content analysis~\cite{oltramari2018privonto, andow2020actions, cui2023poligraph}. 
Among those work, the most important is dataset construction, such as the OPP-115~\cite{wilson2016creation}, 
These datasets have enabled reproducible analyses and cross-domain comparisons, positioning corpora as essential infrastructure for studying transparency practices.

Building on this line of work, we conceptualise AITS as a novel category of regulatory disclosure documents with similar communicative goals but a distinct governance context. 
Thus, we construct one of the first corpus of Government AITS in this paper. 
We believe this corpus and the taxonomy enables systematic stylometric and content analyses of AITS in future, providing empirical insight into how AI transparency is implemented in practice and how effectively regulatory requirements are translated into public-facing disclosures.

\begin{figure}[t]
    \centering
    \includegraphics[width=1\linewidth]{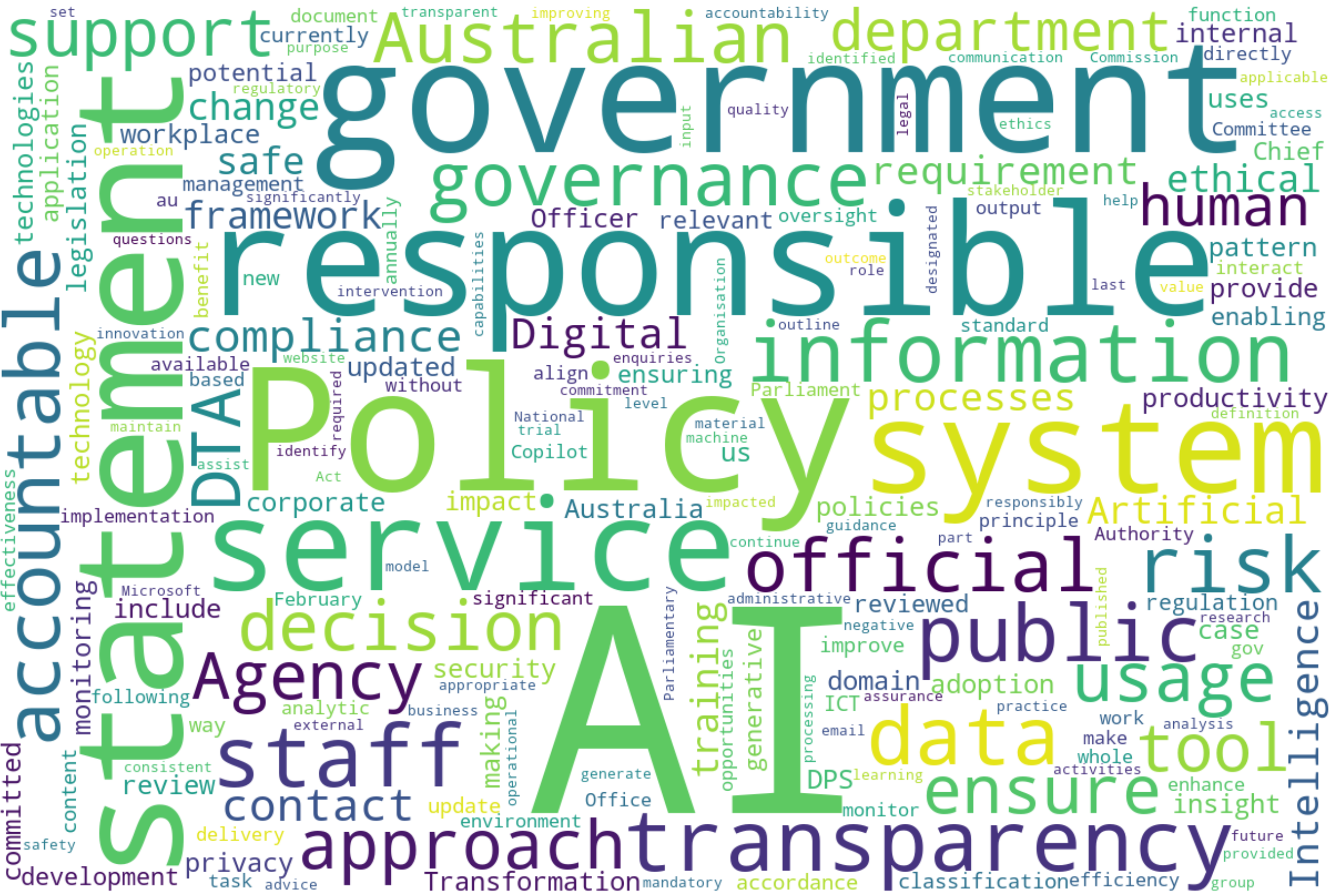}
    \caption{Word cloud of AITS-101.}
    \label{fig:wordcloud}
\end{figure}

\section{Methodology}

\subsection{Scope}

Our study focuses on AITS published by Australian Government entities under the Policy for the Responsible Use of AI in Government. 
Consistent with the policy scope, the target population comprises all non-corporate Commonwealth entities\footnote{\url{https://www.finance.gov.au/about-us/glossary/pgpa/term-non-corporate-commonwealth-entity-nce}} as defined by the Public Governance, Performance and Accountability Act 2013,\footnote{\url{https://www.legislation.gov.au/Details/C2014C00317}} excluding entities that are explicitly exempt under national security or defence-related provisions.
We adopt the official list of NCEs as the authoritative reference for determining entity eligibility. 
Each entity is treated as a single unit of analysis, regardless of internal organisational structure or sub-agencies.

\subsection{Data Collection}

We first compile a comprehensive list of eligible government entities (194) by aggregating portfolio-level listings from official government directories.
Entities falling under defence portfolios or designated as national intelligence agencies were excluded in accordance with the policy exemptions. 
After applying these criteria, the final scope comprises 174 eligible entities across 16 portfolios.

For each eligible entity, we conduct a systematic search for an AITS on the entity’s official public website.\footnote{Their official websites are provided by  Australian government at: \url{https://www.directory.gov.au/portfolios}} 
An AITS is considered valid if 1) It was publicly accessible online without additional authentication; and 2) It was explicitly titled ``\textit{AI Transparency Statement}'' or an unambiguous close variant, for example, ``\textit{Artificial Intelligence Transparency Statement}'' or ``\textit{AI Statement}''.

Search is performed using the website’s internal search function when available, supplemented by manual navigation of policy, governance, and transparency sections. 
Details are available in Appendix~\ref{sec_appendix_searching}.
If no relevant result is returned, then the entity is labelled as no AITS.

We explicitly excluded false positives, including:

\begin{itemize}[noitemsep]
    \item  AI-related project descriptions;
    \item  Academic publications or research programs;
    \item  Privacy policies or accountability statements not specific to AI transparency; and
    \item  General AI news or media releases.
\end{itemize}

At the time of analysis (November 2025), among total of 174 entities, one entity was removed due to administrative restructuring. 
We identify \textbf{101 valid AITS} and confirm 72 entities without an AITS.

\subsection{AITS Preprocessing}
\label{sec:section}

The collected AITS are published in heterogeneous formats, including HTML webpages, PDF files, DOCX documents, and statements embedded within broader policy documents.
HTML pages constituted the majority of statements, while non-HTML formats exhibited substantial variation in layout and structure.
To enable consistent analysis across heterogeneous formats, we convert all AITS into plain text.

We define a section as any heading followed by associated body text. 
Headings include explicit HTML heading tags (e.g., \texttt{<h1>}–\texttt{<h6>}), visually distinct bolded paragraphs, or numbered headings. 
If no heading exists, a visually distinct introductory paragraph is treated as a section with an empty heading.
Nested headings (e.g., ``1'' and ``1.1'') are treated as separate sections when each is followed by distinct body text.

For HTML-based AITS, we apply a rule-based parser to remove non-content elements (e.g., scripts, styles, navigation) and identify headings and associate them with subsequent body text until the next heading. 
We manually check and merge consecutive headings lacking intervening body text.
Minor, we also normalise list items by treating each item as a standalone sentence, appending terminal punctuation when missing.

AITS published as PDF or DOCX files, as well as HTML pages with highly irregular or inconsistent DOM structures, are processed manually. 
We first use the ChatGPT to extract plain text from these documents. 
The extracted content is then manually reviewed to verify that no substantive information was omitted, to ensure correct section segmentation, and to validate sentence boundaries within list items.
Although manual intervention does not reduce processing workload, the relatively small number of such documents makes this approach feasible and ensures consistency and completeness across formats.

\textbf{Dataset.} Table~\ref{tab_dataset_summary} summarises the basic statistics of the dataset, which we refer to as \textbf{AITS-101}. It is available at: \url{https://github.com/ShidongPAN/AITS-101}

\subsection{Annotation Scheme and Process}

We develop an annotation scheme to capture the AI-related practices specified by AITS, following a paradigm similar to that adopted in early privacy policy research~\cite{wilson2016creation}.
To ensure the scheme reflected actual AITS
contents, the annotation scheme is derived through an iterative top-down process, starting from the official AITS requirements and refining categories based on empirical observations from the corpus.
The final annotation scheme consists of ten AI
practice categories:

\begin{itemize} [leftmargin = *]
\item \textbf{General Introduction}: Introductory or contextual content that provides an overview of the AITS, it typically appears at the beginning of the document.

\item \textbf{Definition of AI}: Statements that describe the meaning, scope, or boundaries of AI as used by the agency.

\item \textbf{Intention of AI}: the purposes for which the agency uses or considers adopting AI, capturing organisational motivations and expected benefits.
\item \textbf{AI Usage Patterns}: the classification of AI use according to usage patterns, as defined by the \textit{Classification system for AI use}.
\item \textbf{AI Usage Domains}: the classification of AI use according to usage domains, as defined by the \textit{Classification system for AI use}.
\item \textbf{Impact to the Public}: whether and how AI systems directly or indirectly affect members of the public.
\item \textbf{Monitoring Measures}: Measures to monitor the effectiveness of deployed AI systems and protect the public against negative impacts.
\item \textbf{Compliance with Regulations}: references to compliance with applicable regulations.
\item \textbf{Update}: information about update, such as the publish date or the update frequency. 
\item \textbf{Contact}: contact Information, such as the email address for public enquiries.
\end{itemize}

Similar to~\cite{wilson2016creation}, an individual AI practice belongs to one of
the ten categories above, and it is articulated by
a category-specific set of attributes.
We annotate the ten categories for the total 2,923 sentence.
Notably, as some AITS mentions that they do not use any AI, thus a valid value for each attribute is \textit{no AI use}.
The specific schema is available in the Appendix.

\begin{table}[t]
\centering
  \caption{Basic statistic of AITS-101.}
  \vspace{-5pt}
  \label{tab_dataset_summary}
  \begin{tabular}{l | r}
\toprule
\midrule
    No. AI Transparency Statements & 101 \\
    No. Sections & 688 \\
    No. Sentences & 2923 \\
    \midrule
    Sections per AITS & 6.81 \\
    Sentences per AITS & 28.94 \\
    Words per AITS & 554 \\  
    \midrule
    \bottomrule
\end{tabular}
\end{table}

\section{Stylometric Analysis}

AITS is a recent addition to the landscape of regulatory transparency disclosures, and, to the best of our knowledge, no prior stylometric analyses have been conducted on this document type. Privacy policies serve similar communicative and regulatory functions; therefore, we adopt methodological approaches comsmonly used in stylometric studies of privacy policies \citep{adhikari2023evolution, amos2021privacy, shipp2020private}. 
In addition, we report stylometric analysis results for privacy policies as a reference point for comparison.
For this purpose, we use the Cpp4App dataset \citep{pan2024new, gong2025towards}, a relatively recent collection of privacy policies consisting of 50 popular mobile application privacy policies.

\begin{figure}[t]
    \centering
    \includegraphics[width=0.32\columnwidth]{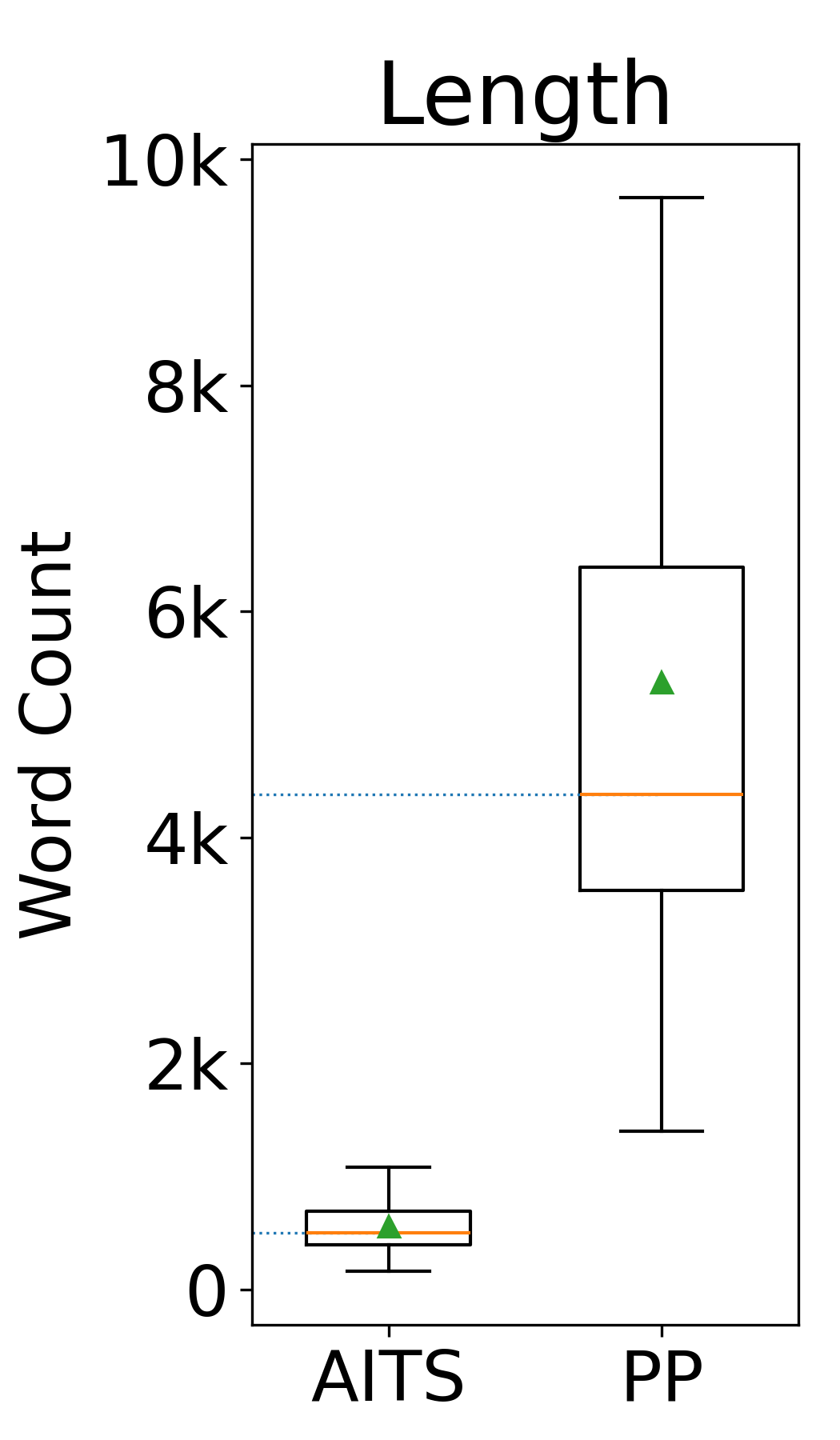}
    \includegraphics[width=0.32\columnwidth]{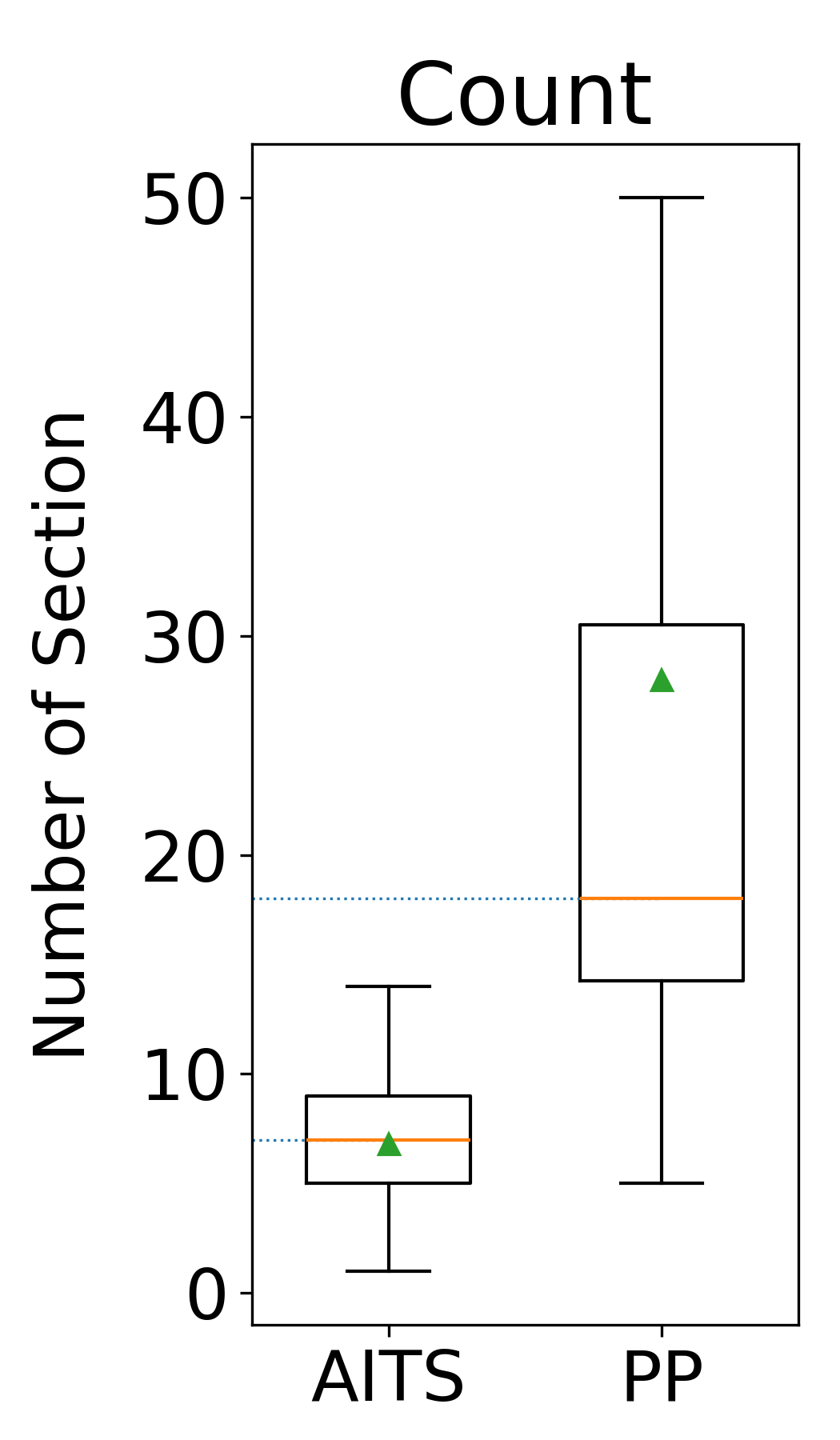}
    \includegraphics[width=0.32\columnwidth]{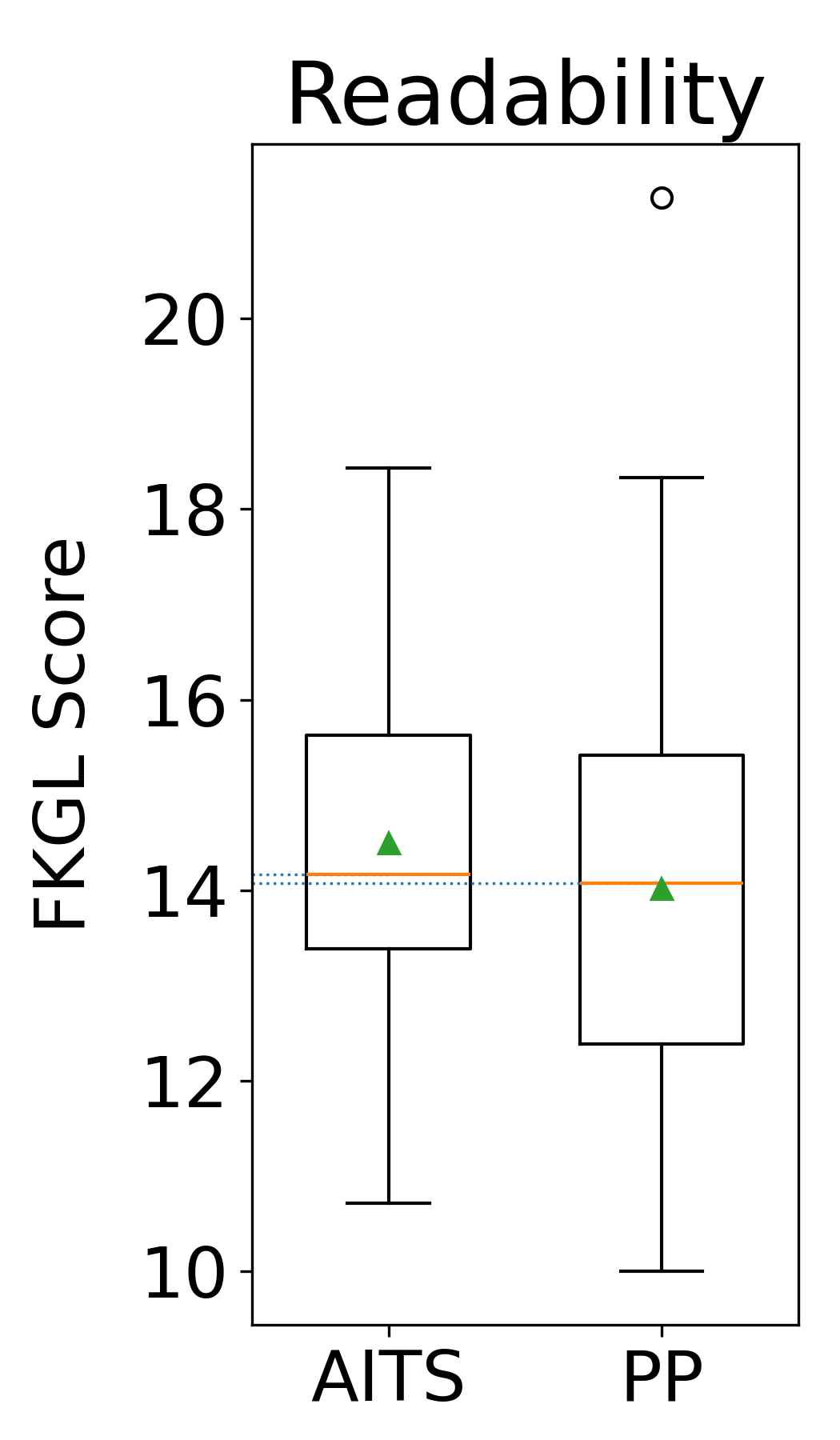}
    \caption{Distributional comparison between the AITS-101 dataset and privacy policies (PP) across three textual characteristics.}
    \label{fig:boxplot}
\end{figure}

\textbf{AITS Length.} 
AITS are not particularly long and are substantially shorter than privacy policies, as shown in Figure~\ref{fig:boxplot}.
In AITS-101, the majority falling between 400 and 700 words, with a median of 498 words and a mean of 554 words. 
By comparison, the median length of privacy policy is 4,379 words, roughly ten times that of AITS. 
This sharp contrast indicates that AITS are far less lengthy and do not exhibit the high degree of textual complexity typically associated with privacy policies.

We further assess AITS length from a structural-complexity perspective by analysing the number of sections in each document, using the sectioning method described in Section~\ref{sec:section}. The AITS-101 dataset shows an average of 6.8 sections per statement. 
Sections are commonly titled as ``\textit{[XX]'s AI adoption and use}'' or ``\textit{Scope and applications}''.
Applying the same segmentation procedure to the privacy policies yields a substantially higher average of 28.0 sections. 
This pronounced difference indicates that AITS are structurally much simpler than privacy policies, reinforcing our earlier observation that AITS are considerably more concise than traditional privacy disclosures.

\textbf{Readability.} 
As a government document intended for public consumption, readability is a central aspect of AITS. 
We assess the readability of all 101 AITSs using the Flesch–Kincaid Grade Level (FKGL) metric \cite{kincaid1975derivation}. 
The median FKGL score is 14.16, indicating that a reader generally requires at least college-level proficiency to comfortably understand the content—well above the level easily accessible to the general public. 
This score is nearly identical to the median FKGL of privacy policies in the privacy policies dataset (14.07), indicating that despite being substantially shorter in length, AITS retain the same readability limitations commonly associated with traditional privacy policies. 
Overall, these findings indicate that AITS remain challenging for the general public to comprehend, reflecting a substantial barrier to effective public access and understanding.

\begin{table}[t]
\centering
\caption{Four linguistic dimensions extracted from the plain language guidance.}
\label{tab_plain_language}

\begingroup
\hyphenpenalty=10000
\exhyphenpenalty=10000

\begin{tabular}{
    >{\centering\arraybackslash}p{0.05\linewidth} |
    >{\raggedright\arraybackslash}p{0.34\linewidth} |
    >{\raggedright\arraybackslash}p{0.47\linewidth}
}
\toprule
\textbf{ID} & \textbf{Dimension} & \textbf{Description} \\
\midrule
D1 & Simple Words and Expressions &
Use everyday words and simple expressions. \\
\midrule
D2 & Jargon and Shortened-Form &
Avoidance of jargon; appropriate use of shortened forms, with explanation. \\
\midrule
D3 & Use of Personal Pronouns &
Appropriate use of we, you, and your to address readers directly. \\
\midrule
D4 & Inclusive Language &
Respectful and inclusive wording acknowledging diverse individuals and backgrounds. \\
\bottomrule
\end{tabular}

\endgroup
\end{table}

\textbf{Lexical Features.} The Australian Government’s Standard for AITS specifically requires agencies to ``\textit{use clear, plain language}'' in accordance with the guidance provided in the Plain language and word choice section within the Australian Government Style Manual~\cite{australian-gov-style-manual-plain-language}. 
Motivated by this requirement, we conduct a lexical analysis of AITS to assess whether their linguistic characteristics align with the Standard’s expectations.

Based on the guidance, we abstract four dimensions that capture relevant requirements as the basis for our analysis, as shown in Table~\ref{tab_plain_language}:    
\textbf{(1)} Simple Words and Expressions. 
\textbf{(2)} Jargon and Shortened-Form. 
\textbf{(3)} Use of Personal Pronouns. 
\textbf{(4)} Inclusive Language. 
To evaluate how well ACTS comply with these four dimensions in large-scale, we employ an MLLM-based annotation approach followed by human validation.
Specifically, we use GPT-5 to batch-assess all AITS-101 documents, assigning 1 point for each dimension that was met and 0 otherwise, resulting in a maximum possible score of 4 per AITS, and requiring the model to provide a justification for each judgement to support subsequent validation.
The complete prompt is available in the Appendix.
Afterward, we randomly sample ten annotated instances and asked an experienced researcher in responsible AI and privacy policy to manually annotate. 
The Cohen’s kappa \cite{cohen1960coefficient} between human and the GPT-5 is 0.70, which indicates substantial agreement. 
Therefore, we adopt the LLM-generated annotations in the following analysis.

Analysis of the annotation results shows that the 101 AITSs have an average total score of 1.55 out of 4, indicating that overall compliance with the four plain-language dimensions is limited. 
Most AITS perform poorly on \textit{Simple Words and Expressions}. 
Specifically, none of the AITS received a point for Simple Words and Expressions.
Closer examination reveals several recurring issues, including the pervasive use of bureaucratic and abstract wording (e.g., \textit{``assess'', ``prepare'', ``engage'', ``utilise''}) and syntactically complex sentences (e.g., \textit{``The utilisation and introduction of AI into its operations is guided by... ''}).
Performance on \textit{Jargon and Shortened-Form} is similarly weak: only 5\% of AITS score positively on this dimension. 
Common problems include the use of specialist jargon, such as \textit{``data-driven decision augmentation''} and \textit{``human-in-the-loop''}, that is unlikely to be readily understood by the general public, as well as the introduction of uncommon abbreviations or technical terms without explanation, for example, the frequent use of \textit{``ICT'}' without expanding it as Information and Communication Technology.

On the other hand, more than half of the AITSs (52\%) appropriately use \textit{Personal Pronouns} to directly engage the readers. 
Moreover, 100\% of the AITS satisfyingly follow the guidance for \textit{Inclusive Language}. 
Overall, while AITS exhibit strong adherence to inclusive language norms, their reliance on bureaucratic vocabulary, abstract phrasing, and insufficient control of jargon suggests that they remain challenging for the public to easily understand.

To provide a comparison baseline, we collect a comparison sample of other 50 government documents, including privacy policies, copyright statements, and accessibility statements, and other government-issued public-facing documents. They are randomly drawn from the same Australian department websites from which the AITS are sourced. 
We apply the same evaluation procedure to this corpus and the results broadly mirror those of the AITS.
The mean score for the comparative documents is 1.84, with dimension-level compliance rates of 6\%, 14\%, 64\%, and 100\% respectively.
This pattern suggests that bureaucratic and abstract wording are common features of Australian government documents in general, whereas inclusive language is consistently well implemented. Nonetheless, AITS appear to be slightly more difficult for the public to understand compared to other government documents.

\begin{figure}[t]
    \centering
    \begin{subfigure}[t]{0.49\columnwidth}
        \centering
        \includegraphics[width=\linewidth]{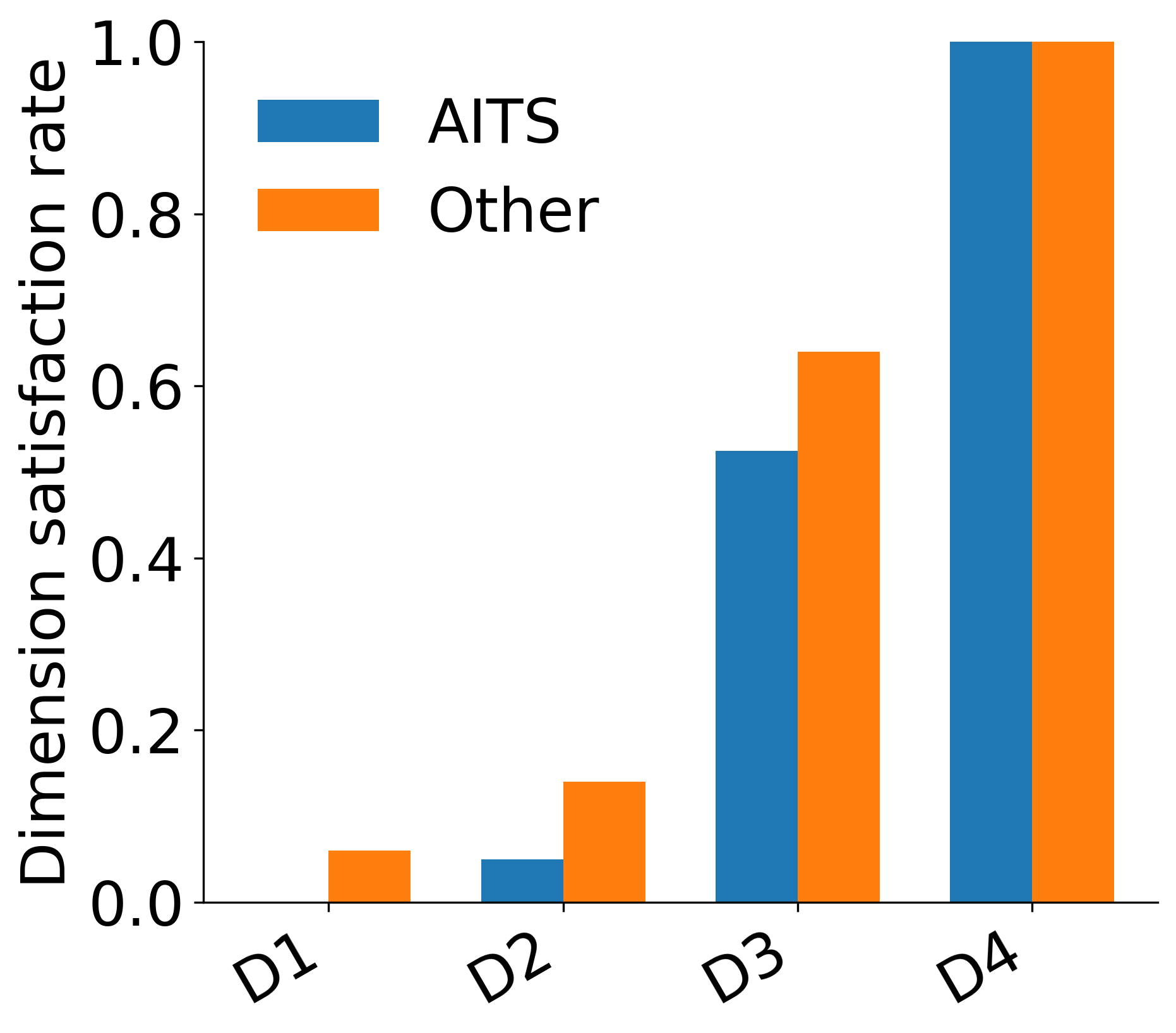}
        \caption{Dimension satisfaction.}
        \label{fig:lexical_bar_grouped}
    \end{subfigure}
    \hfill
    \begin{subfigure}[t]{0.49\columnwidth}
        \centering
        \includegraphics[width=\linewidth]{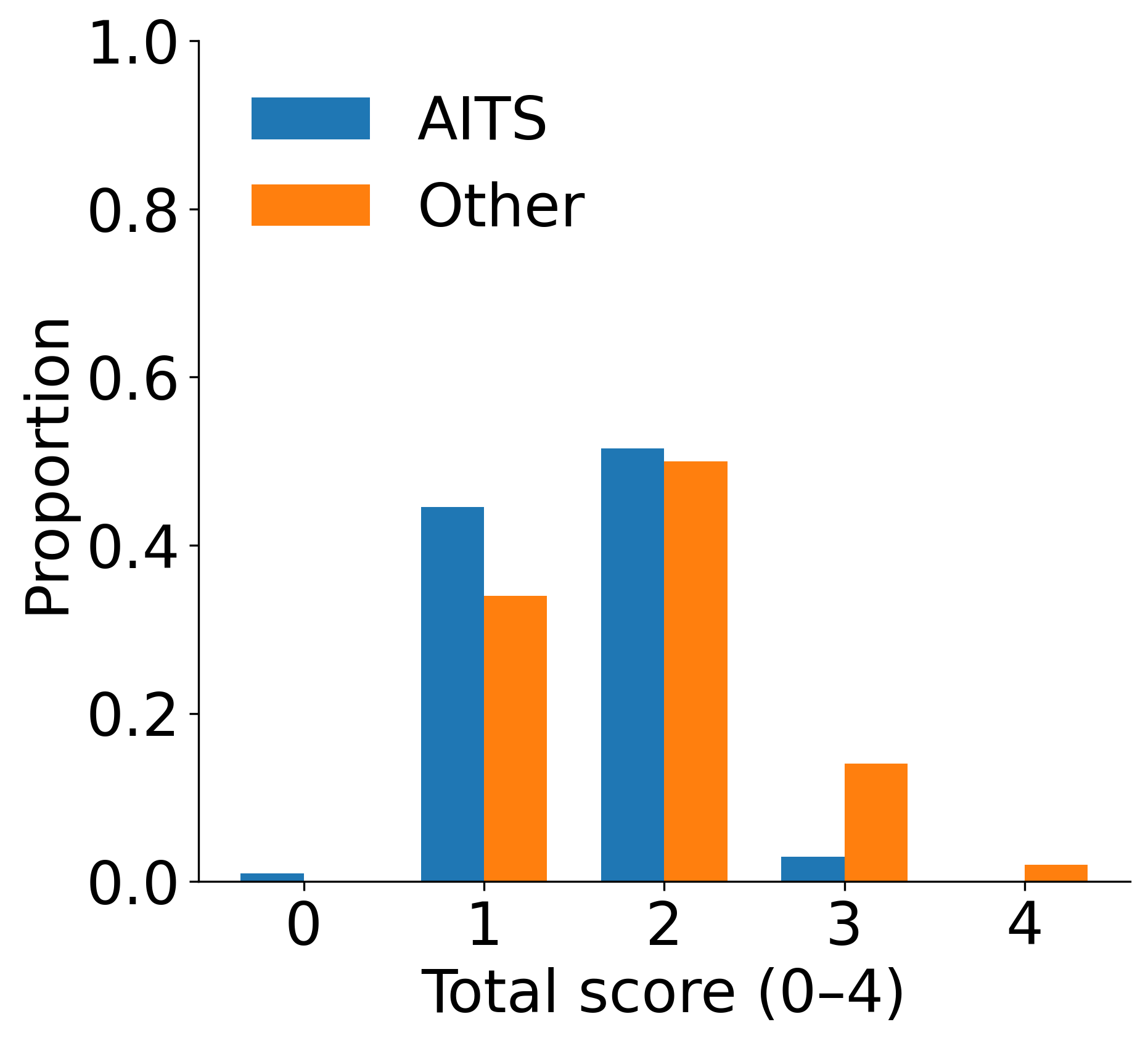}
        \caption{Total score distribution.}
        \label{fig:total_score_grouped}
    \end{subfigure}
\vspace{-5pt}
    \caption{Comparison of lexical features analysis results between AITS-101 and other government documents.}
    \label{fig:plain_language_comparison}
    \vspace{-5pt}
\end{figure}









\begin{table*}[t]
\centering
\caption{By-category descriptive statistics for annotation classes in the AITS corpus. 
Frequency denotes the total number of annotated segments per category. 
Mean and median are calculated across all AITS. 
Coverage represents the proportion of AITS containing at least one annotation of the corresponding category.}
\label{tab:l1-composition}
\small
\begin{tabular}{l|r|r|r||r}

\textbf{Category} & \textbf{Freqency} & \textbf{Mean} & \textbf{Median} & \textbf{Coverage} \\
\hline
Compliance with Regulations    & 829 & 8.2 & 7 & 100\% \\
Monitoring Measures            & 365 & 3.6 & 3 & 92.08\% \\
AI Usage Patterns              & 338 & 3.4 & 3 & 91.09\% \\
Update                         & 300 & 3.0 & 3 & 98.02\% \\
AI Usage Domains               & 285 & 2.8 & 2 & 80.20\% \\
\hline
Contact                        & 176 & 1.7 & 1 & 94.06\%\\
Intention of AI                & 176 & 1.7 & 1 & 79.28\% \\
Impact to the Public           & 165 & 1.6 & 1 & 74.26\% \\
General Introduction           & 118 & 1.2 & 1 & 58.41\% \\
Definition of AI               & 68 & 0.7 & 0 & 40.59\% \\
\end{tabular}
\vspace{-5pt}
\end{table*}

\section{Composition of AITS-101 Dataset}

The AITS corpus enables an examination of how government agencies structure their AITS in terms of disclosed AI-related practices. 
Although AITS are generally concise compared to privacy policies, the brevity does not necessarily imply comprehensive coverage of transparency requirements. 
Table~\ref{tab:l1-composition} summarises the pre-consolidation counts (\textit{i.e.,} frequency) of annotated segments across the AI practice categories, together with the mean and median number of segments per AITS. 
Overall, the composition of AITS is characterised by a strong emphasis on governance-facing disclosures rather than descriptive explanations of AI systems.
In particular, Compliance with Regulations dominates the corpus both in frequency and coverage, appearing in all AITS with a mean of 8.2 annotated segments per statement.
This reflects the policy-driven nature of AITS and suggests that agencies prioritised adherence to mandated requirements.

Monitoring Measures appear in over 92\% of statements, indicating that most agencies provide at least some information about how AI use is overseen, governed, or reviewed. 
This high coverage suggests that agencies recognise monitoring as a core expectation of AI transparency.
Additionally, Update information also shows high coverage at 98.02\%, suggesting that metadata related to statement maintenance is widely included.
Moreover, AI Usage Patterns are reported in more than 91\%, and content describing AI Usage Domains and Intentions of AI is less consistently reported.

In contrast, definitions of AI are provided in only 40.59\% of AITS, with a median of zero annotated segments per statement. This indicates that many agencies assume a shared or implicit understanding of what constitutes AI, rather than explicitly defining its scope or boundaries. 
However, as discussed in Section~\ref{subsec:definitions}, the absence of clear definitions can lead to ambiguity in how AI use is framed and understood, particularly when agencies adopt narrow or inconsistent interpretations of AI systems.

Similar to~\citep{wilson2016creation}, \textit{coverage} in Table~\ref{tab:l1-composition} is defined in an ipso facto sense: an AI-related practice category is considered to cover a policy segment if the corresponding theme can be interpreted from the segment text.
Notably, a statement with a higher number of annotated segments may reflect verbosity or redundancy, while a statement with fewer annotations may simply be concise or narrowly scoped. 
In both cases, the presence or absence of specific disclosure categories provides insight into how agencies operationalise transparency requirements, rather than serving as a direct measure of accountability or risk mitigation.

\section{Patterns and Implications}
\label{sec:patterns}
This section identifies four structural patterns that characterise the Australian Government's approach to AI transparency. 
We examine how the legibility of government automation is mediated by the machinery of government itself--specifically through technical definitions, shared service dependencies, and procurement standardisation. These patterns illustrate how agencies navigate the tension between administrative efficiency and requirements to signal risk in an evolving technological environment.

\subsection{Definitional Scope: Syntax versus Provenance}
\label{subsec:definitions}

A foundational characteristic of the transparency corpus is the reliance on strict technical definitions to delineate the scope of reporting. 
Agencies consistently apply the OECD requirement~\cite{OECD} for ``inference'' to distinguish novel AI applications from established automated systems.
Consequently, the corpus reflects a categorical exclusion of ``rules-based'' systems or deterministic Automated Decision Making. 
For instance, the Australian Taxation Office articulates:
\vspace{-5pt}
\begin{quote}
    \textit{``We don't consider rules-based analytics...to be AI. This form of analytics does not infer how to generate outputs from the inputs they receive.''} \cite{ato_ai_transparency_statement}
\end{quote}
\vspace{-5pt}
While this exclusion aligns with standards, it relies on a \textit{syntactic} distinction (whether the system uses probabilistic inference or deterministic logic) rather than an assessment of the system's \textit{provenance}. By grouping all deterministic systems into a single ``non-AI'' category, the current framework treats \textit{expert-authored rules} (logic explicitly coded by humans) and \textit{statistically induced rules} (logic learnt from data but operationalised deterministically) as functionally equivalent.

This focus on the form of the system rather than its origin establishes a specific boundary for risk assessment. It allows for a scenario where data-derived models, once frozen into static rules, fall outside the transparency reporting requirements while potentially retaining the underlying model risks inherent in their training data. An expanded transparency regime might distinguish systems based on the validation and origin of their logic, rather than solely on the syntax of their execution.

\subsection{The ``High-Stakes Exclusion Zone'': Negative Assurance}
\label{subsec:highstakes}

For agencies operating within high-sensitivity domains, the transparency statements function primarily as instruments of ``negative assurance'': defining the agency's AI posture by what it is \textit{not} doing. We observe a distinct trend where agencies with direct responsibility for vulnerable populations present the most restrictive definitions of AI usage.

Agencies such as Services Australia~\cite{servicesaustralia_ai_transparency_statement}, the Department of Veterans' Affairs~\cite{dva_ai_transparency_statement}, and the Administrative Review Tribunal~\cite{art_ai_transparency_statement} employ categorical exclusions regarding autonomous decision-making. Services Australia, for example, qualifies its schema entry for ``Decision Making'' with a definitive boundary:
\begin{quote}
    \textit{``We do \textbf{not} use AI to process claims... final decisions or actions are made by a human.''} \cite{servicesaustralia_ai_transparency_statement}
\end{quote}

This reliance on the ``human-in-the-loop'' safeguard is ubiquitous across the social services sector. However, while the statements confirm the \textit{presence} of human oversight, they generally lack detail regarding the \textit{mechanisms} of that oversight (e.g., review protocols or intervention rates). This creates a transparency gap where the human safeguard is asserted as a binary status rather than an operational process.

A notable divergence from this conservative posture is found in the Australian Bureau of Statistics, which explicitly discloses a public-facing generative AI chatbot for the Census Test~\cite{abs_ai_transparency_statement}. 
This contrast highlights a sector-specific variance in risk appetite: while statistical agencies appear empowered to pilot generative interfaces, agencies with adjudicative or welfare responsibilities maintain a strict delineation between automation and decision-making, likely reflecting the higher stakes of error in administrative justice.

\subsection{Structural Dependency: The Governance of Scale versus Context}
\label{subsec:dependency}

The corpus reflects the extent to which transparency reporting is shaped by the operational realities of the machinery of government. We observe a recurrent ``Parent-Child'' governance model, where smaller statutory agencies leverage the digital infrastructure of larger portfolio departments. This arrangement prioritises the scale and consistency of centralized IT governance, ensuring that smaller entities benefit from the security frameworks and procurement power of larger departments.

This dynamic is explicitly codified within the Parliamentary Portfolio. The Department of the Senate~\cite{aph_senate_ai_transparency_statement}, theDepartment of the House of Representatives~\cite{aph_dhr_ai_transparency_statement_2025}, and the Parliamentary Budget Office~\cite{pbo_ai_transparency_statement_2025} formally note their reliance on the Department of Parliamentary Services~\cite{aph_dps_ai_transparency_statement_2025} for ICT provision. Similarly, independent bodies in the Health portfolio operate as tenants within the Department of Health and Aged Care's digital environment.

While this centralized model maximises operational efficiency, it creates a distributed form of transparency. The resulting disclosures often reference the ``Parent'' entity's broad governance policies, which aggregates the risk reporting at the portfolio level. Consequently, while context-specific risk assessments may be conducted internally by the ``Child'' agency, the public-facing transparency statements effectively decouple the agency's specific use cases from the infrastructure governing them. This requires the public to navigate a composite reporting structure, cross-referencing agency declarations with portfolio-level disclosures.

\subsection{The ``Copilot'' Standardisation: Platform Assurance versus Operational Risk}
\label{subsec:copilot}

Quantitatively, the most frequently reported use case across the corpus is ``Workplace Productivity,'' a category dominated by a high degree of vendor concentration. A significant proportion of agencies, including the policy-setting Digital Transformation Agency~\cite{dta_ai_transparency_statement_2025}, list Microsoft 365 Copilot as a primary AI application.
This widespread adoption indicates that the initial phase of generative AI is being operationalised through established procurement channels. By deploying AI as a feature set within existing licensing agreements, agencies effectively leverage the safety and alignment frameworks of a major global vendor to manage \textit{platform risk}, such as data security and model integrity.

However, while procurement controls manage the platform, they do not inherently address the \textit{downstream task delegation risk}. The integration of generative capabilities into standard office suites lowers the friction for high-stakes use cases, such as drafting policy or summarizing sensitive briefs, which effectively expands the agency's risk surface. This suggests a phenomenon of \textit{capability diffusion}: as AI becomes a standard utility, the primary challenge shifts from vetting the technology itself to managing the distributed, everyday decisions of the staff employing it.

\subsection{Implications} 
Collectively, these patterns illustrate that the current transparency regime is optimised for \textit{organisational accountability} and \textit{risk signalling} rather than technical granularity. By successfully identifying governance hierarchies and demarcating the boundaries of agency activity, the framework fulfills its primary administrative function. The distinction between this governance-focused reporting and \textit{system-level technical transparency} suggests that the AITS framework is currently architected to authorise administrative action. Consequently, the variation between public interest in technical details and the administrative focus of the statements reflects a difference in audience expectations, where documents designed for governance assurance are viewed through the lens of technical auditability. This is not a failure of the AITS regime, but a reminder that \textbf{organisational transparency cannot substitute for system-level assurance mechanisms}.

\section{Conclusion}

This paper provides one of the first systematic empirical examination of AI Transparency Statements (AITS) and a government AITS dataset, dubbed AITS-101.
Using stylometric, quantitative, and qualitative document analyses, we examine disclosure coverage, structure, and recurring patterns. 
Our findings reveal substantial variation in AI-related practice disclosure, highlight gaps between policy intent and implementation, and inform the design of more effective public-sector AI transparency standards.

\section*{Limitations}

Despite the measures taken to ensure the reliability of our analysis, several limitations should be acknowledged. 
First, the lexical features analysis relies on GPT-5 to assess the use of plain language in AITS. 
Although we conducted manual spot checks and observed substantial agreement between human judgment and GPT-5 assessments, the model occasionally produces misclassifications or deviates from the intended prompt criteria. 
In particular, GPT-5 may lack the contextual and pragmatic understanding required to fully account for established conventions in government and legal writing. 
For example, GPT-5 may interpret abbreviated forms appearing in the formal names of legislation or institutions as unexplained shortened-form under the Jargon and Shortened-Form dimension, whereas such terms function as institutionally standardised proper names. This reflects a tendency toward surface-level pattern matching rather than nuanced interpretation of domain-specific context. Future work could examine the robustness of lexical feature assessments by comparing results across different large language models or evaluation settings.

In addition, our use of binary (0/1) scoring for lexical dimensions introduces a degree of simplification. While this design choice improves consistency and interpretability, it may obscure subtle variations in language quality. For instance, under the Use of Personal Pronouns dimension, some statements partially adopt personal pronouns while still frequently referring to the responsible department by name; such mixed usage cannot be adequately captured by a binary outcome and may therefore underrepresent intermediate or transitional writing styles. More fine-grained or graded scoring schemes could be explored in future work to better capture variations in language quality.

Second, the sentence-level annotation of compositional elements in the AITS-101 dataset is subject to inherent subjectivity. Although the annotators engaged in extensive discussion and alignment of category definitions to improve consistency, judgment may still diverge in complex or ambiguous cases. As a result, some degree of annotation noise is unavoidable, and the applied mitigation strategies reduce but do not fully eliminate this limitation. Future research may benefit from larger annotator pools or complementary automated methods to further mitigate subjectivity in annotation.

\bibliography{custom}

@article{czarnitzki2023artificial,
  title={Artificial intelligence and firm-level productivity},
  author={Czarnitzki, Dirk and Fern{\'a}ndez, Gast{\'o}n P and Rammer, Christian},
  journal={Journal of Economic Behavior \& Organization},
  volume={211},
  pages={188--205},
  year={2023},
  doi={10.1016/j.jebo.2023.05.008}
}

@article{wirtz2019artificial,
  title={Artificial intelligence and the public sector—applications and challenges},
  author={Wirtz, Bernd W and Weyerer, Jan C and Geyer, Carolin},
  journal={International Journal of Public Administration},
  volume={42},
  number={7},
  pages={596--615},
  year={2019},
  publisher={Taylor \& Francis}
}

@article{mergel2024implementing,
  title={Implementing AI in the public sector},
  author={Mergel, Ines and Dickinson, Helen and Stenvall, Jari and Gasco, Mila},
  journal={Public Management Review},
  pages={1--14},
  year={2024},
  publisher={Taylor \& Francis}
}

@article{brynjolfsson2025generative,
  title={Generative AI at work},
  author={Brynjolfsson, Erik and Li, Danielle and Raymond, Lindsey},
  journal={The Quarterly Journal of Economics},
  volume={140},
  number={2},
  pages={889--942},
  year={2025},
  publisher={Oxford University Press}
}

@misc{australia-ai-transparency-standard-2025,
  key          = {Digital Transformation Agency (Australia)},
  author       = {{Digital Transformation Agency, Australian Government}},
  title        = {Standard for AI Transparency Statements},
  year         = {2025},
  howpublished = {\url{https://www.digital.gov.au/ai/ai-in-government-policy/standard-ai-transparency-statements}},
  url          = {https://www.digital.gov.au/ai/ai-in-government-policy/standard-ai-transparency-statements},
  note         = {Supporting the Policy for the Responsible Use of AI in Government; last updated 1 December 2025},
  urldate      = {2025-12-11}
}

@article{adhikari2023evolution,
  title={Evolution of composition, readability, and structure of privacy policies over two decades},
  author={Adhikari, Andrick and Das, Sanchari and Dewri, Rinku},
  journal={Proceedings on Privacy Enhancing Technologies},
  year={2023}
}

@inproceedings{amos2021privacy,
  title={Privacy policies over time: Curation and analysis of a million-document dataset},
  author={Amos, Ryan and Acar, Gunes and Lucherini, Eli and Kshirsagar, Mihir and Narayanan, Arvind and Mayer, Jonathan},
  booktitle={Proceedings of the Web Conference 2021},
  pages={2165--2176},
  year={2021}
}

@article{shipp2020private,
  title={How private is your period?: A systematic analysis of menstrual app privacy policies},
  author={Shipp, Laura and Blasco, Jorge},
  journal={Proceedings on Privacy Enhancing Technologies},
  year={2020}
}

@inproceedings{pan2024new,
  title={A $\{$NEW$\}$$\{$HOPE$\}$: Contextual Privacy Policies for Mobile Applications and An Approach Toward Automated Generation},
  author={Pan, Shidong and Tao, Zhen and Hoang, Thong and Zhang, Dawen and Li, Tianshi and Xing, Zhenchang and Xu, Xiwei and Staples, Mark and Rakotoarivelo, Thierry and Lo, David},
  booktitle={33rd USENIX Security Symposium (USENIX Security 24)},
  pages={5699--5716},
  year={2024}
}

@techreport{kincaid1975derivation,
  title={Derivation of new readability formulas (automated readability index, fog count and flesch reading ease formula) for navy enlisted personnel},
  author={Kincaid, J Peter and Fishburne Jr, Robert P and Rogers, Richard L and Chissom, Brad S},
  year={1975}
}

@article{tao2025longitudinal,
  title={A Longitudinal Measurement of Privacy Policy Evolution for Large Language Models},
  author={Tao, Zhen and Pan, Shidong and Xing, Zhenchang and Black, Emily and Gillis, Talia B and Chen, Chunyang},
  journal={arXiv preprint arXiv:2511.21758},
  year={2025},
  url={https://arxiv.org/abs/2511.21758}
}

@article{lu2024responsible,
  title={Responsible AI pattern catalogue: A collection of best practices for AI governance and engineering},
  author={Lu, Qinghua and Zhu, Liming and Xu, Xiwei and Whittle, Jon and Zowghi, Didar and Jacquet, Aurelie},
  journal={ACM Computing Surveys},
  volume={56},
  number={7},
  pages={1--35},
  year={2024},
  publisher={ACM New York, NY}
}

@misc{ato_ai_transparency_statement,
  author       = {{Australian Taxation Office}},
  title        = {ATO AI transparency statement},
  year         = {2025},
  url          = {https://www.ato.gov.au/about-ato/commitments-and-reporting/information-and-privacy/ato-ai-transparency-statement}
}

@inproceedings{pan2024trap,
  title={Is it a trap? a large-scale empirical study and comprehensive assessment of online automated privacy policy generators for mobile apps},
  author={Pan, Shidong and Zhang, Dawen and Staples, Mark and Xing, Zhenchang and Chen, Jieshan and Xu, Xiwei and Hoang, Thong},
  booktitle={33rd USENIX Security Symposium (USENIX Security 24)},
  pages={5681--5698},
  year={2024}
}

@article{chang2018role,
  title={The role of privacy policy on consumers’ perceived privacy},
  author={Chang, Younghoon and Wong, Siew Fan and Libaque-Saenz, Christian Fernando and Lee, Hwansoo},
  journal={Government Information Quarterly},
  volume={35},
  number={3},
  pages={445--459},
  year={2018},
  publisher={Elsevier}
}

@inproceedings{andow2020actions,
  title={Actions speak louder than words:$\{$Entity-Sensitive$\}$ privacy policy and data flow analysis with $\{$PoliCheck$\}$},
  author={Andow, Benjamin and Mahmud, Samin Yaseer and Whitaker, Justin and Enck, William and Reaves, Bradley and Singh, Kapil and Egelman, Serge},
  booktitle={29th USENIX Security Symposium (USENIX Security 20)},
  pages={985--1002},
  year={2020}
}

@inproceedings{cui2023poligraph,
  title={$\{$PoliGraph$\}$: Automated privacy policy analysis using knowledge graphs},
  author={Cui, Hao and Trimananda, Rahmadi and Markopoulou, Athina and Jordan, Scott},
  booktitle={32nd USENIX Security Symposium (USENIX Security 23)},
  pages={1037--1054},
  year={2023}
}

@article{schaub2025grand,
  title={Grand Challenges for Research on Privacy Documents},
  author={Schaub, Florian and Utz, Christine and Wilson, Shomir and Xian, Lu},
  year={2025}
}

@article{oltramari2018privonto,
  title={PrivOnto: a semantic framework for the analysis of privacy policies},
  author={Oltramari, Alessandro and Piraviperumal, Dhivya and Schaub, Florian and Wilson, Shomir and Cherivirala, Sushain and Norton, Thomas B and Russell, N Cameron and Story, Peter and Reidenberg, Joel and Sadeh, Norman},
  journal={Semantic Web},
  volume={9},
  number={2},
  pages={185--203},
  year={2018},
  publisher={SAGE Publications Sage UK: London, England}
}

@inproceedings{wilson2016creation,
  title={The creation and analysis of a website privacy policy corpus},
  author={Wilson, Shomir and Schaub, Florian and Dara, Aswarth Abhilash and Liu, Frederick and Cherivirala, Sushain and Leon, Pedro Giovanni and Andersen, Mads Schaarup and Zimmeck, Sebastian and Sathyendra, Kanthashree Mysore and Russell, N Cameron and others},
  booktitle={Proceedings of the 54th Annual Meeting of the Association for Computational Linguistics (Volume 1: Long Papers)},
  pages={1330--1340},
  year={2016}
}

@misc{servicesaustralia_ai_transparency_statement,
  author       = {{Services Australia}},
  title        = {Automation and Artificial Intelligence Transparency Statement},
  year         = {2025},
  url          = {https://www.servicesaustralia.gov.au/automation-and-artificial-intelligence-transparency-statement}
}

@article{olsen2024right,
  title={The right to transparency in public governance: Freedom of information and the use of artificial intelligence by public agencies},
  author={Olsen, Henrik Palmer and Hildebrandt, Thomas Troels and Wiesener, Cornelius and Larsen, Matthias Smed and Fl{\"u}gge, Asbj{\o}rn William Ammitzb{\o}ll},
  journal={Digital Government: Research and Practice},
  volume={5},
  number={1},
  pages={1--15},
  year={2024},
  publisher={ACM New York, NY}
}

@misc{abs_ai_transparency_statement,
  author       = {{Australian Bureau of Statistics}},
  title        = {AI transparency statement},
  year         = {2025},
  url          = {https://www.abs.gov.au/about/legislation-and-policy/ai-transparency-statement}
}

@misc{dva_ai_transparency_statement,
  author       = {{Department of Veterans’ Affairs}},
  title        = {Artificial Intelligence (AI) Transparency Statement},
  year         = {2025},
  url          = {https://www.dva.gov.au/about-us/corporate-governance/artificial-intelligence-ai-transparency-statement}
}

@misc{art_ai_transparency_statement,
  author       = {{Administrative Review Tribunal}},
  title        = {AI transparency statement},
  year         = {2025},
  url          = {https://www.art.gov.au/ai-transparency-statement}
}

@misc{aph_senate_ai_transparency_statement,
  author       = {{Parliament of Australia, Department of the Senate}},
  title        = {AI transparency statement},
  year         = {2025},
  url          = {https://www.aph.gov.au/About\_Parliament/Parliamentary\_departments/Department\_of\_the\_Senate/Accountability\_and\_reporting/AI\_transparency\_statement}
}

@misc{aph_dhr_ai_transparency_statement_2025,
  author       = {{Parliament of Australia, Department of the House of Representatives}},
  title        = {AI Transparency Statement},
  year         = {2025},
  url          = {https://www.aph.gov.au/-/media/05_About_Parliament/54_Parliamentary_Depts/542_Dept_of_House_of_Reps/Documents/DHR_AI_Transparency_Statement_-_Final_1.pdf}
}

@misc{pbo_ai_transparency_statement_2025,
  author       = {{Parliamentary Budget Office}},
  title        = {AI Transparency Statement},
  year         = {2025},
  url          = {https://www.pbo.gov.au/about-the-pbo/performance-and-reporting/information-and-privacy/ai-transparency-statement}}

@misc{aph_dps_ai_transparency_statement_2025,
  author       = {{Parliament of Australia, Department of Parliamentary Services}},
  title        = {AI Transparency Statement},
  year         = {2025},
  url          = {https://www.aph.gov.au/About_Parliament/Parliamentary_departments/Department_of_Parliamentary_Services/Publications/AI_Transparency_Statement}
}

@article{matulionyte2023regulating,
  title={Regulating transparency of AI: a survey of best practices},
  author={Matulionyte, Rita},
  journal={Available at SSRN 4554868},
  year={2023}
}

@misc{dta_ai_transparency_statement_2025,
  author       = {{Digital Transformation Agency}},
  title        = {AI transparency statement},
  year         = {2025},
  url          = {https://www.dta.gov.au/ai-transparency-statement}
}

@misc{OECD,
  author       = {{Organisation for Economic Co-operation and Developmen}},
  title        = {OECD Artificial intelligence},
  year         = {2025},
  url          = {https://www.oecd.org/en/topics/policy-issues/artificial-intelligence.html}
}

@misc{australian-gov-style-manual-plain-language,
  title        = {Plain language and word choice},
  author       = {{Australian Government Style Manual}},
  year         = {2025},
  url          = {https://www.stylemanual.gov.au/writing-and-designing-content/clear-language-and-writing-style/plain-language-and-word-choice},
  note         = {Accessed: 2025-12-31},
}

@article{andrews2022trust,
  title={A Trust Framework for Government Use of Artificial Intelligence and Automated Decision Making},
  author={Andrews, Pia and de Sousa, Tim and Haefele, Bruce and Beard, Matt and Wigan, Marcus and Palia, Abhinav and Reid, Kathy and Narayan, Saket and Dumitru, Morgan and Morrison, Alex and others},
  journal={arXiv preprint arXiv:2208.10087},
  year={2022}
}

@article{lund2025standards,
  title={Standards, frameworks, and legislation for artificial intelligence (AI) transparency},
  author={Lund, Brady and Orhan, Zeynep and Mannuru, Nishith Reddy and Bevara, Ravi Varma Kumar and Porter, Brett and Vinaih, Meka Kasi and Bhaskara, Padmapadanand},
  journal={AI and Ethics},
  pages={1--17},
  year={2025},
  publisher={Springer}
}

@article{sebastiao2025ai,
  title={AI Transparency: A Conceptual, Normative, and Practical Frame Analysis},
  author={Sebasti{\~a}o, S{\'o}nia Pedro and Dias, David Ferreira-Mendes},
  journal={Media and Communication},
  volume={13},
  year={2025}
}

@article{cohen1960coefficient,
  title={A coefficient of agreement for nominal scales},
  author={Cohen, Jacob},
  journal={Educational and psychological measurement},
  volume={20},
  number={1},
  pages={37--46},
  year={1960},
  publisher={Sage Publications Sage CA: Thousand Oaks, CA}
}

@article{pan2023toward,
  title={Toward the cure of privacy policy reading phobia: Automated generation of privacy nutrition labels from privacy policies},
  author={Pan, Shidong and Hoang, Thong and Zhang, Dawen and Xing, Zhenchang and Xu, Xiwei and Lu, Qinghua and Staples, Mark},
  journal={arXiv preprint arXiv:2306.10923},
  year={2023}
}

@article{hoang2026esg,
  title={ESG Reporting Lifecycle Management with Large Language Models and AI Agents},
  author={Hoang, Thong and Klymenko, Mykhailo and Xu, Xiwei and Pan, Shidong and Ding, Yi and Tang, Xushuo and Yang, Zhengyi and Shi, Jieke and Lo, David},
  journal={arXiv preprint arXiv:2603.10646},
  year={2026}
}

@article{zhang2025right,
  title={Right to be forgotten in the era of large language models: Implications, challenges, and solutions},
  author={Zhang, Dawen and Finckenberg-Broman, Pamela and Hoang, Thong and Pan, Shidong and Xing, Zhenchang and Staples, Mark and Xu, Xiwei},
  journal={AI and Ethics},
  volume={5},
  number={3},
  pages={2445--2454},
  year={2025},
  publisher={Springer}
}

@article{gong2025towards,
  title={Towards Context-aware Mobile Privacy Notice: Implementation of A Deployable Contextual Privacy Policies Generator},
  author={Gong, Haochen and Tao, Zhen and Pan, Shidong and Xing, Zhenchang and Sun, Xiaoyu},
  journal={arXiv preprint arXiv:2509.22900},
  year={2025}
}

@report{weatherall2026aitransparency,
  author      = {Weatherall, Kimberlee and Bello y Villarino, Jos{\'e}-Miguel and Sinclair, Abbe},
  title       = {AI Transparency in Practice: An Evaluation of Commonwealth Entities' Compliance with Their Obligations Regarding AI Transparency Statements},
  institution = {ARC Centre of Excellence for Automated Decision Making and Society},
  address     = {Melbourne},
  year        = {2026},
  doi         = {10.60836/gbx1-8w63}
}

\appendix

\section{Searching AITSs on Official Websites}
\label{sec_appendix_searching}

This appendix describes the procedure used to identify and collect AITS from Australian Government entities.

We first obtained a list of Australian Government departments and agencies from the official government directory at \url{https://www.directory.gov.au/portfolios}. For each listed entity, we attempted to access its official website via the provided link. In cases where a website was not listed or the link was unavailable or non-functional, we supplemented this process by searching for the entity’s official website using Google.

Once an entity’s official website was identified, we followed a structured search strategy to locate its AITS. If the website provided an internal search function, we prioritised its use and queried the term ``AI Transparency Statement''. 
If this search returned a webpage clearly titled by AI Transparency Statement for the entity, we treated it as a valid AITS and included it in our dataset. If the search returned results but none were related to an AI Transparency Statement, or if no results were returned at all, the entity was marked as No AITS.

For websites without an internal search function, we conducted manual navigation of sections commonly used to host policy and governance-related content, including (but not limited to) policy, governance, transparency, and about pages. If no webpage explicitly presenting an AI Transparency Statement could be identified through this process, the entity was also marked as No AITS.

This procedure was applied consistently across all entities to ensure comparability and reproducibility of the data collection process.

\section{AITS Content Composition Taxonomy}
\onecolumn

\setlength{\tabcolsep}{4pt}
\renewcommand{\arraystretch}{1.15}
\footnotesize

\begin{longtable}{
c
P{\dimexpr0.18\linewidth\relax}
P{\dimexpr0.24\linewidth\relax}
P{\dimexpr0.18\linewidth\relax}
P{\dimexpr0.34\linewidth\relax}
}
\caption{Classification Framework for AITS's Content}
\label{tab:aits_classification_framework} \\

\toprule
\textbf{ID} &
\textbf{Level 1 Class} &
\textbf{Definition} &
\textbf{Level 2 Class} &
\textbf{Definition} \\
\midrule
\endfirsthead

\toprule
\textbf{ID} &
\textbf{L1 Class} &
\textbf{Definition} &
\textbf{L2 Class} &
\textbf{Definition} \\
\midrule
\endhead

\midrule
\multicolumn{5}{r}{\textit{Continued on next page}} \\
\endfoot

\bottomrule
\endlastfoot

\multirow{1}{*}{1} &
\multirow{1}{=}{\textbf{Definition of AI}} &
\multirow{1}{=}{\emph{-}} &
\textbf{Definition of AI} &
\emph{Statements that describe the meaning, scope, or boundaries of AI as used by the agency. It includes formal definitions of AI, explanations of what constitutes an AI system or AI tool, and illustrative examples provided to clarify what the agency considers to be AI within the context of the transparency statement.} \\
\midrule

\multirow{1}{*}{2} &
\multirow{1}{=}{\textbf{General introduction}} &
\multirow{1}{=}{\emph{-}} &
\textbf{General introduction} &
\emph{Introductory or contextual content that provides an overview of the AI Transparency Statement. It typically appears at the beginning of the document and outlines the purpose, scope, or background of the statement without describing specific AI systems or practices in detail.} \\
\midrule

\multirow[t]{12}{*}{3} &
\multirow[t]{12}{=}{\textbf{Intention of AI}} &
\multirow[t]{12}{=}{\emph{The intentions behind why the agency uses AI or is considering its adoption.}} &
\textbf{Productivity} &
\emph{AI is intended to improve internal efficiency, reduce administrative workload, or streamline organisational processes, including administration efficiency, team communication, and paperwork processing.} \\
\cline{4-5}
& & & \textbf{Service delivery} & \emph{AI is intended to enhance the delivery of services to the public, such as improving timeliness, accessibility, quality, or responsiveness of government services.} \\
\cline{4-5}
& & & \textbf{Decision making} & \emph{AI is intended to support, inform, or augment decision-making processes, including risk assessment, prioritisation, recommendations, or eligibility evaluations, with decisions made by human officials.} \\
\cline{4-5}
& & & \textbf{Innovation} & \emph{AI is intended to enable new capabilities, modernise existing systems, or support novel approaches beyond routine efficiency gains.} \\
\cline{4-5}
& & & \textbf{Staff experience} & \emph{AI is intended to improve staff working experience, such as reducing cognitive burden, supporting learning, or enhancing workplace support and collaboration.} \\
\cline{4-5}
& & & \textbf{Research} & \emph{AI is intended for research, experimentation, or analytical studies, including pilot projects or exploratory investigations without full operational deployment.} \\
\cline{4-5}
& & & \textbf{Exploration of potential} & \emph{Exploration about feasibility, benefits, risks, or limitations of AI technologies, often in early-stage trials or proofs of concept.} \\
\cline{4-5}
& & & \textbf{Support regulatory} & \emph{AI is intended to support regulatory functions, such as monitoring compliance, enforcing regulations, or assisting regulatory oversight activities.} \\
\cline{4-5}
& & & \textbf{Communication} & \emph{AI is intended to support internal and external communication activities.} \\
\cline{4-5}
& & & \textbf{Data processing and analysis} & \emph{AI is intended to process, analyse, or extract insights from data, including document analysis, data analysis, classification, or pattern detection.} \\
\cline{4-5}
& & & \textbf{Auditing} & \emph{AI is intended to support auditing activities.} \\
\cline{4-5}
& & & \textbf{No AI use} & \emph{Specifically mention that there is no AI usage.} \\
\midrule

\multirow[t]{5}{*}{4} &
\multirow[t]{5}{=}{\textbf{AI usage patterns}} &
\multirow[t]{5}{=}{\emph{The classification of AI use according to usage patterns, as defined by the Attachment A - Classification system for AI use.}} &
\textbf{Decisionmaking and administrativeaction} &
\emph{Used to either: support decision making or the taking ofadministrative action by guiding, assessing, or makinga recommendation to a human decision makermake decisions or take administrative action withouthuman intervention.Note: not all automated decision making may beconsidered AI (noting the definition under the policy).} \\
\cline{4-5}
& & & \textbf{Analytics for insights} & \emph{Identifies, produces or understands insightswithin structured or unstructured materials viacomprehensive data analysis, predictive modellingand/or reporting tools.} \\
\cline{4-5}
& & & \textbf{Workplace productivity} & \emph{Automates routine tasks, manage workflows, andfacilitate communication.} \\
\cline{4-5}
& & & \textbf{Image processing} & \emph{Processes images to automatically identify patternsand objects, such as faces, for official purposes.} \\
\cline{4-5}
& & & \textbf{No AI use} & \emph{Specifically mention that there is no AI usage.} \\
\midrule

\multirow[t]{7}{*}{5} &
\multirow[t]{7}{=}{\textbf{AI usage domains}} &
\multirow[t]{7}{=}{\emph{Classification of AI use according to usage domains, as defined by the Attachment A - Classification system for AI use.}} &
\textbf{Service deliver} &
\emph{Enhances efficiency or accuracy of government services, including paymentservices, by providing tailored and responsive services to the public.This may include in direct interaction with the public, such as chat-bots,enhanced customer self-service and multilingual capabilities, or support staff orsystems which deliver services.} \\
\cline{4-5}
& & & \textbf{Compliance and fraud detection} & \emph{Identifies patterns or anomalies in data to detect fraudulent activities and ensurecompliance with laws and regulations.} \\
\cline{4-5}
& & & \textbf{Law enforcement, intelligence and security} & \emph{Supports law enforcement and intelligence agencies by analysing data fromvarious sources to predict and prevent crimes, and by aiding in intelligencegathering.} \\
\cline{4-5}
& & & \textbf{Policy and legal} & \emph{Analyses policies and legal documents to provide advice and assurance on theirimpact and supports policy development that is consistent with existing laws.} \\
\cline{4-5}
& & & \textbf{Scientific} & \emph{Leveraged in scientific endeavours to process complex datasets, simulateexperiments, predict outcomes and enhance monitoring functions.} \\
\cline{4-5}
& & & \textbf{Corporate and enabling} & \emph{Supports corporate functions, including HR, finance, media and communications,and IT, by automating processes, optimising resource allocation and improvingoperational efficiency.} \\
\cline{4-5}
& & & \textbf{No AI use} & \emph{Specifically mention that there is no AI usage.} \\
\midrule

\multirow[t]{3}{*}{6} &
\multirow[t]{3}{=}{\textbf{Impact to the public}} &
\multirow[t]{3}{=}{\emph{Classification of use where the public may directly interact with, or be significantly impacted by, AI or its outputs without human review.}} &
\textbf{Direct public interaction} &
\emph{Public directly interact with or get exposed to AI systems or receive AI-generated outputs.} \\
\cline{4-5}
& & & \textbf{Indirect public interaction} & \emph{AI systems might influence decisions, processes, or outcomes that affect the public, but without direct interaction.} \\
\cline{4-5}
& & & \textbf{No AI use} & \emph{Specifically mention that there is no AI usage.} \\
\midrule

\pagebreak[4]
\multirow[t]{10}{*}{7} &
\multirow[t]{10}{=}{\textbf{Monitoring measures}} &
\multirow[t]{10}{=}{\emph{Measures to monitor the effectiveness of deployed AI systems and protect the public against negative impacts.}} &
\textbf{Management} &
\emph{Governed under dedicated management, governance, or risk-management frameworks, including formal oversight structures or assurance processes.} \\
\cline{4-5}
& & & \textbf{Internal policy} & \emph{Internal organisational policies, guidelines, or standards are established to regulate the responsible development, deployment, or use of AI systems.} \\
\cline{4-5}
& & & \textbf{Internal register} & \emph{AI usage are documented in internal inventories or registers that track their existence, purpose, or risk status.} \\
\cline{4-5}
& & & \textbf{Staff training} & \emph{Training or capability-building activities are provided to staff involved in AI development, use, oversight, or decision-making.} \\
\cline{4-5}
& & & \textbf{Collaboration across departments} & \emph{Monitoring and oversight involve collaboration, coordination, or transparency between different departments, teams, or organisational units.} \\
\cline{4-5}
& & & \textbf{Accountable officials} & \emph{Specific roles, positions, or officers are formally designated as accountable for AI governance, oversight, or compliance.} \\
\cline{4-5}
& & & \textbf{Specific legal frameworks} & \emph{AI use is subject to explicit legal or regulatory frameworks, beyond general internal policies.} \\
\cline{4-5}
& & & \textbf{Policy review and evaluate} & \emph{AI systems, related policies, or governance arrangements are periodically reviewed, evaluated, or updated to assess effectiveness and manage risks.} \\
\cline{4-5}
& & & \textbf{Manual validation on generated content} & \emph{AI-generated outputs are subject to human review, validation, or approval before being used, acted upon, or communicated externally.} \\
\cline{4-5}
& & & \textbf{No AI use} & \emph{Specifically mention that there is no AI usage.} \\
\midrule

\multirow[t]{2}{*}{8} &
\multirow[t]{2}{=}{\textbf{Compliance with AI regulations and policies}} &
\multirow[t]{2}{=}{\emph{Mentions of specific compliance or applicable legislation and regulations.}} &
\textbf{"The Policy for responsible use of AI in government"} &
\emph{Compliance with "the Policy for responsible use of AI in government".} \\
\cline{4-5}
& & & \textbf{Other legislations and regulations} & \emph{SCompliance with other legislations and regulations.} \\
\midrule

\multirow{1}{*}{9} &
\multirow{1}{=}{\textbf{Update}} &
\multirow{1}{=}{\emph{-}} &
\textbf{Update} &
\emph{Statements about the AITS update, such as the publish date or the update frequency.} \\
\midrule

\multirow{1}{*}{10} &
\multirow{1}{=}{\textbf{Contact}} &
\multirow{1}{=}{\emph{-}} &
\textbf{Contact} &
\emph{Contact Information, such as the email address for public enquiries.} \\
\midrule

\end{longtable}

\section{Prompt for Lexical Features Analysis}

\begin{tcolorbox}[
  colback=white,
  colframe=black,
  boxrule=0.6pt,
  arc=0mm,
  left=6pt,
  right=6pt,
  top=6pt,
  bottom=6pt,
]

\# Evaluator Instructions\\
You are a strict evaluator.\\
Evaluate the given text against the Australian Government Style Manual:  
\emph{Plain language and word choice}  
\url{https://www.stylemanual.gov.au/writing-and-designing-content/clear-language-and-writing-style/plain-language-and-word-choice}
Return ONLY valid JSON.

\medskip
\# Scoring rules\\
Be conservative: give 1 only when clearly satisfied overall.
Judge appropriateness within government communication, not casual writing.

\medskip
\# Dimensions (binary 0/1)\\
1) simple\_word\_choice \\
1 if the text mostly uses everyday words and phrases acceptable in government writing, prefers simple words over complicated expressions, and avoids or only minimally uses bureaucratic or formulaic language. \\
0 otherwise.
\medskip\\
2) jargon\_and\_shortened\_form\_control \\
1 if jargon, slang, and idioms are avoided; widely recognised shortened forms (e.g., DNA) may be used; less common abbreviations/acronyms are only used for recurring concepts and are explained at first use. \\
0 otherwise.
\medskip\\
3) personal\_pronouns \\
1 if, where appropriate, the text mainly uses personal pronouns (we/you/us/your) to speak directly. \\
0 otherwise.
\medskip\\
4) inclusive\_language \\
1 if wording is generally respectful and inclusive, respecting all people, their rights, and their heritage. \\
0 otherwise.

\medskip
\# Output JSON schema (exact keys)
\begin{lstlisting}[style=jsonwrap]
{
  "simple_word_choice": 0 or 1,
  "simple_word_choice_reason": "Explain the reason for the score; if 0, explain the problem.",
  "simple_word_choice_example": "If 0, cite one example sentence; otherwise, return an empty string.",

  "jargon_and_shortened_form_control": 0 or 1,
  "jargon_and_shortened_form_control_reason": "Explain the reason for the score; if 0, explain the problem.",
  "jargon_and_shortened_form_control_example": "If 0, cite one example sentence; otherwise, return an empty string.",

  "personal_pronouns": 0 or 1,
  "personal_pronouns_reason": "Explain the reason for the score; if 0, explain the problem.",
  "personal_pronouns_example": "If 0, cite one example sentence; otherwise, return an empty string.",

  "inclusive_language": 0 or 1,
  "inclusive_language_reason": "Explain the reason for the score; if 0, explain the problem.",
  "inclusive_language_example": "If 0, cite one example sentence; otherwise, return an empty string."
}
\end{lstlisting}

\end{tcolorbox}

\end{document}